\documentclass[12pt]{article}
\usepackage{amsmath}
\usepackage[pdftex]{graphicx}
\usepackage{psfrag,epsf}
\usepackage{enumerate}
\usepackage[dvipsnames]{xcolor}
\usepackage{natbib}
\usepackage{url} 
\usepackage{soul}

\usepackage{rotating}
\usepackage{makecell}

\usepackage{amsfonts} 

\usepackage{array}
\usepackage{multirow}

\usepackage{tabularx}

\usepackage[skip=0.5pt]{caption} 

\newcommand{\blind}{0}

\begin{document}

\def\spacingset#1{\renewcommand{\baselinestretch}%
{#1}\small\normalsize} \spacingset{1}


\if0\blind
{
  \title{\bf Boldness-Recalibration for Binary Event Predictions}
  \author{Adeline P. Guthrie \\
    Department of Statistics, Virginia Tech\\
    and \\
    Christopher T. Franck\thanks{
    Corresponding author}\hspace{.2cm}\\
    Department of Statistics, Virginia Tech}
  \maketitle
} \fi

\if1\blind
{
  \bigskip
  \bigskip
  \bigskip
  \begin{center}
    {\LARGE\bf Title}
\end{center}
  \medskip
} \fi

\bigskip
\begin{abstract}
Probability predictions are essential to inform decision making across many fields. Ideally, probability predictions are (i) well calibrated, (ii) accurate, and (iii) bold, i.e., spread out enough to be informative for decision making. However, there is a fundamental tension between calibration and boldness, since calibration metrics can be high when predictions are overly cautious, i.e., non-bold.  The purpose of this work is to develop a Bayesian model selection-based approach to assess calibration, and a strategy for boldness-recalibration that enables practitioners to responsibly embolden predictions subject to their required level of calibration. Specifically, we allow the user to pre-specify their desired posterior probability of calibration, then maximally embolden predictions subject to this constraint. We demonstrate the method with a case study on hockey home team win probabilities and then verify the performance of our procedures via simulation. We find that very slight relaxation of calibration probability (e.g., from 0.99 to 0.95) can often substantially embolden predictions when they are well calibrated and accurate (e.g., widening hockey predictions’ range from 26\%-78\% to 10\%-91\%).
\end{abstract}

\noindent%
\vfill

\newpage
\spacingset{1.45} 
\section{Introduction}
\label{sec:intro}

Probability predictions are made for everyday events, from the mundane, like the probability it will rain, to the life-altering, like the probability that a natural disaster hits a particular city.  These predictions arise from both sophisticated statistical and machine learning techniques and/or simply from human judgement and expertise. Regardless, probability predictions are commonly used in important decision making processes in the fields of medicine, economics, image recognition via machine learning, sports analytics, entertainment and many others, so it is critical that we have methods that assess such predictions. The purpose of this paper is to develop boldness-recalibration that enables forecasters to achieve well calibrated, responsibly bold probability predictions for binary events.  

We describe the assessment of probability predictions in terms of three aspects: calibration, accuracy, and boldness. The first aspect is calibration.  Predicted probabilities are well calibrated when the events they aimed to predict occur with the same probability that was forecasted. For example, if the home team wins 40$\%$ of games for which a hockey forecaster predicted a 40$\%$ chance of a win, the forecaster is well calibrated.  Probability calibration is well studied within the fields of statistics, meteorology, psychology, machine learning, and others \citep{Bross1953,MurphyWinkler1977,Dawid1982,DeGroot1983,Gonzalez1999, Guoetal2017}. Naturally, calibration is considered a minimal desirable property of predicted probabilities \citep{Dawid1982}.  Without calibration, if a forecaster says the probability of a home team win is 70$\%$, you cannot rely on that prediction to reflect the true probability of a win. However, well calibrated predictions are not necessarily accurate nor bold enough to be useful. 

The second aspect to assess probability predictions is classification accuracy.  Classification accuracy measures how well predictions distinguish between the events they aim to predict.  Receiver-Operating Characteristic (ROC) curves and the corresponding area under the ROC curve (AUC) are frequently used to assess classification accuracy for probability predictions.  These accuracy assessments do not measure calibration since any monotone transformation applied to forecaster probability predictions will produce the same ROC curve and AUC as the original predictions. 

The third aspect to assess probability predictions is boldness. We define boldness simply as the spread (i.e. standard deviation) in probability predictions.  To illustrate, in National Hockey League (NHL) games in the 2020-21 season, the home team won 53$\%$ of games, thus the sample proportion is $\bar y = $ 0.53.  A hockey forecaster who simply predicts the base rate of 0.53 for every NHL game is well calibrated, but lacks the boldness needed for actionable predictions.  However, forecasters who produce bold predictions alone without calibration or good classification show they have misplaced confidence in their prediction ability.  

\begin{figure}[!ht]
\begin{center}
\includegraphics[width=2in]{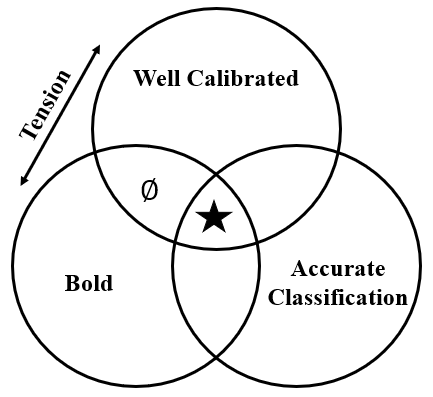}
\end{center}
\caption{Venn Diagram highlighting the possible combinations of three aspects of probability predictions: calibration, boldness, and classification accuracy. We propose a boldness-recalibration approach that enables forecasters to maximize boldness while maintaining a high probability of calibration, subject to their classification accuracy. The star represents probability predictions that are well calibrated, bold, and accurate. Empty set $\phi$ indicates that forecasts that are not accurate cannot be both well calibrated and bold.\label{fig:venn_diag}}
\end{figure}

This brings into focus the core tension between calibration and boldness, subject to classification accuracy. Notice the star in Figure \ref{fig:venn_diag} at the center of the three aspects.  This intersection is where forecaster predictions are accurate, well calibrated, reasonably bold, and thus, actionable for decision making. It is important to note that the level of boldness considered ``reasonable'' depends directly on classification accuracy and the overall decision making goal.  In some cases (such as forecasting rain), maintaining the highest level of calibration achievable may be most important. Increasing boldness here may be useful so that event planners can consider well informed ``worst case" weather predictions for a specific day. In other settings, poor classification accuracy (e.g., due to badly informed forecasters)  may limit the amount of emboldening that is responsible. Sacrificing a small amount of calibration for greater boldness allows the analyst to responsibly examine riskier predictions in a variety of areas where investments of time, effort, and/or money are called for (e.g., sports betting, medical diagnostics, financial investing, hiring employees, etc.). This paper allows the analyst to examine emboldened probability predictions in the context of a user-specified requirement for calibration using Bayesian reasoning.

Several techniques exist that purely focus on correcting miscalibration. \cite{Dalton2013} leverages the Cox linear-logistic model to test for calibration and proposes a relative calibration metric, but makes no mention of prediction boldness.  \cite{Platt2000} introduces Platt scaling which recalibrates Support Vector Machine (SVM) output via sigmoid curves. \cite{Guoetal2017} proposes recalibration via temperature scaling, a one-parameter extension of Platt scaling for Neural Network output. \cite{ZadroznyElkan2001} propose a non-parametric approach called histogram binning where probabilities are bin-wise recalibrated to minimize squared loss.  \cite{ZadroznyElkan2002} extends this by fitting a piece-wise recalibration function on each bin interval.  \cite{NaeiniCooperHauskrecht2015} extends this further to Bayesian Binning into Quantiles (BBQ) where multiple binning strategies are considered via Bayesian model averaging. However, none of these methods incorporate boldness in their adjustment.

Some approaches to assessing calibration also consider notions similar to boldness. Reliability diagrams plot the predicted forecaster probabilities versus the observed frequency within each bin \citep{MurphyWinkler1977}.   A calibration metric called Expected Calibration Error (ECE) quantifies the miscalibration seen in reliability diagrams by averaging the distances between the predictions and observed frequencies within each bin.  Sometimes histograms \citep{RanjanGneiting2010, Dimitriadis2021} or density plots \citep{Satopaa2022} of the predicted probabilities are included with reliability diagrams to visualize boldness.  However, boldness is not quantified in this approach.

A common metric for prediction accuracy, the Brier Score, can be decomposed into three parts such that one component  measures calibration, another measures resolution, and the last measures the uncertainty in the outcomes themselves \citep{Brier1950, Murphy1973a}.  Resolution (or discrimination) refers to how well forecasts distinguish between the two possible outcomes.  Predicted probabilities have high resolution when they are further from the base rate.  While resolution is similar in concept to boldness, they are not mathematically equivalent.  We will show that boldness is measured independently of the event outcomes, where measures of resolution rely on the base rate, and thus are not fully disentangled from the overall uncertainty of event outcomes.

A few methods both recalibrate and embolden predictions in highly specific circumstances.  \cite{Lichtendahletal2022} and \cite{Satopaa2022} focus on aggregates of forecaster predictions and their spread but they specifically focus on a forecast aggregation approach that is not applicable to individual forecasters. We focus on appropriately emboldening predictions from a single forecaster subject to their calibration and classification accuracy.  The predictions in the case study we present are not aggregate forecasts, and thus the approaches of \cite{Lichtendahletal2022} and \\cite{Satopaa2022} are not applicable here.  \cite{Turner2014}, \cite{Baronetal2014}, and \cite{Atanasovetal2015} also focus on aggregates, using the linear log odds (LLO) recalibration function to adjust aggregate boldness.  \cite{Roitbergetal2022} employs a network based temperature scaling approach to recalibrate and correct overly bold softmax pseudo-probabilities.  However, \cite{Turner2014}, \cite{Baronetal2014}, \cite{Atanasovetal2015} and \cite{Roitbergetal2022} all rely on the Brier Score and/or ECE to assess calibration. \cite{HanBudescu2022} focus on LLO applied to forecasts of continuous, rather than binary, events. \cite{Gonzalez1999} use LLO to recalibrate single forecaster predictions but focus solely on the psychological implications of probability perception for binary events.  None of these methods provide direct control of the calibration-boldness tradeoff. 

To the best of our knowledge, no methodology yet exists that provides a mechanism to directly control the tradeoff between calibration and boldness. To address this gap in the literature, we propose boldness-recalibration.  Boldness-recalibration allows users to set the desired level of calibration in terms of the posterior calibration probability and then maximizes boldness by maximizing spread in predictions subject to calibration level.  Three key virtues of this approach are that it (a) quantifies the calibration-boldness tradeoff in an interpretable manner (in a Bayesian sense), (b) is forecaster agnostic, meaning it operates only on probability and event data, not on how the forecaster made the predictions, and (c) does not rely on binning. 

The rest of this paper is organized as follows.  Section \ref{sec:meth} introduces boldness-recalibration methodology and pertinent real-world and simulation data examples.  In Section \ref{sec:res}, we provide the results of our real-world and simulated examples.  Section \ref{sec:conc} provides a discussion and concluding comments.

\section{Methods}
\label{sec:meth}

The following approaches are forecaster agnostic, meaning they can be applied to any probability forecasts of binary events produced by forecasters from many domains, regardless of how the predictions were made.   By forecaster, we mean any entity that produces probability predictions, regardless if they are machine learning output and/or a product of human judgement and expertise. 

\subsection{Linear Log Odds (LLO) Recalibration Function}
\label{sec:LLO}

To assess calibration, we use the linear log odds (LLO) recalibration function. Let $c(x_i;\delta, \gamma)$ be the LLO function 
\begin{equation}\label{llo}
    c(x_i;\delta, \gamma) = \frac{\delta x_i^\gamma}{\delta x_i^\gamma + (1-x_i)^\gamma},
\end{equation}
where $x_i$ is a probability prediction from a forecaster, $\delta > 0$ and $\gamma \in \mathbb{R}$.  We call the outputted probability, $c(x_i;\delta, \gamma)$, the LLO-adjusted probability.  The LLO-adjusted set is based on shifting and scaling each of the original forecaster probabilities $x_i$ on the log odds scale using $\delta$ and $\gamma$.  Thus, on the log odds scale, the LLO-adjusted set is linear with respect to $x_i$ according to intercept $log(\delta)$ and slope $\gamma$, and can be re-written as 
\begin{equation} \label{LinRel}
   log\left(\frac{c(x_i;\delta, \gamma)}{1-c(x_i;\delta, \gamma)}\right) = \gamma log\left(\frac{x_i}{1-x_i}\right)+log(\delta).
\end{equation}
Suitable choices of $\delta$ and $\gamma$ can calibrate poorly calibrated probabilities. The flexibility of the LLO function can capture many forms of miscalibration \citep{Gonzalez1999, Turner2014}.  When both $\delta = 1$ and $\gamma = 1$, the LLO function imposes no shifting nor scaling, returning the original prediction $x_i$ \citep{Gonzalez1999}.  Thus, null values of $\delta_0 = \gamma_0 = 1$ corresponds to the hypothesis that $x_i$ is well calibrated. This is similar to how Reliability Diagrams operate in that when predicted forecaster probabilities are close to the observed frequency within each bin (i.e. forecasts are well calibrated), the result resembles the x=y line.  The same is true when plotting event rates by LLO-adjusted probability forecasts via $\delta = 1$ and $\gamma = 1$ under calibration. It is important to note that if $c(x_i;\delta, \gamma)$ is considered the LLO-adjusted ``event'' probability, the corresponding LLO-adjusted ``non-event'' probability is $1 - c(x_i;\delta, \gamma)$ rather than $c(1 - x_i;\delta, \gamma)$.

\subsection{Likelihood Function}
\label{sec:likelihood}

We adopt a Bernoulli likelihood where the events are presumed independent and the probability of each event is governed by LLO-adjusted probabilities. Let $\mathbf{y}$ be a vector of $n$ outcomes corresponding to the $n$ predictions in $\mathbf{x}$ from a single forecaster.  Then, we have
\begin{equation}\label{likelihood}
    \pi(\mathbf{x}, \mathbf{y} | \delta, \gamma) = \prod_{i=1}^n c(x_i;\delta, \gamma)^{y_i} \left[1-c(x_i;\delta, \gamma)\right]^{1-y_i}.
\end{equation}
This likelihood enables calibration maximization via maximum likelihood estimates (MLEs) for $\delta$ and $\gamma$.  The $\hat\delta_{MLE}$ and $\hat\gamma_{MLE}$ values produce optimally calibrated probabilities, $c(x_i; \hat\delta_{MLE}, \hat\gamma_{MLE})$.  Shifting via $\hat\delta_{MLE}$ on the log odds scale adjusts the average prediction to match the sample proportion.  Scaling by $\hat\gamma_{MLE}$ on the log odds spreads out or contracts predictions based on accuracy.  This may be a desirable approach when probability calibration is the sole priority.  Our approach of adopting a Bernoulli Likelihood governed by LLO-adjusted probabilities is equivalent to a specialized logistic regression model.  Details can be found in the online supplement.

\subsection{Bayesian Assessment of Calibration}
\label{sec:bt}

Using the likelihood function in the previous section, we take a Bayesian model selection-based approach to calibration assessment.  We compare a  well calibrated model, $M_c$ (where $\delta=\gamma=1$), to an uncalibrated model, $M_{u}$ (where $\delta>0, \gamma \in \mathbb{R}$).  The posterior model probability of  $M_c$ given the observed outcomes $\mathbf{y}$ serves as our measure of calibration for the testing framework and can be expressed as
\begin{align} \label{postM1}
P(M_c|\mathbf{y}) = \frac{P(\mathbf{y}|M_c) P(M_c)}{P(\mathbf{y}|M_c) P(M_c) + P(\mathbf{y}|M_{u}) P(M_{u})}.
\end{align}
Here, $P(\mathbf{y}|M_i)$ is the integrated likelihood of the observed outcomes $\mathbf{y}$ given $M_i$ and $P(M_i)$ is the prior probability of model $i$, $i \in \{c,u\}$. The Bayes Factor comparing the uncalibrated model to the calibrated model is defined as
\begin{align}
    BF = \frac{P(\mathbf{y}|M_{u})}{P(\mathbf{y}|M_c)}.
\end{align}
Inverting (\ref{postM1}) gives us
\begin{align}
\frac{1}{P(M_c|\mathbf{y})} &= \frac{P(\mathbf{y}|M_c) P(M_c)}{P(\mathbf{y}|M_c) P(M_c)} + \frac{P(\mathbf{y}|M_u) P(M_u)}{P(\mathbf{y}|M_c) P(M_c)}\\
 &= 1 + BF \frac{P(M_u)}{P(M_c)}
\end{align}
Thus the expression in (\ref{postM1}) can be re-written as
\begin{align} \label{pcalib}
    P(M_c|\mathbf{y}) = \frac{1}{1 + BF \frac{P(M_u)}{P(M_c)}}.
\end{align}

An essential component of Bayesian model selection is the specification of prior model probabilities. To the best of our knowledge, this is the first attempt to assess calibration probability through Bayesian model selection, and thus best practices for setting $P(M_c)$ and $P(M_u)$ have not yet been established. We set these prior probabilities to $\frac{1}{2}$ in subsequent analyses for illustrative purposes only.  Fully reproducible code will be made available with final acceptance of the article in the online supplement, which will allow users to set alternate model priors.

Using the likelihood in (\ref{likelihood}), the integrated likelihoods, $P(\mathbf{y}|M_i)$, are not analytically tractable.  While a fully Bayesian approach could be implemented, we advocate for a useful approximation. We employ a large sample approximation to the Bayes factor using the Bayesian Information Criteria (BIC) such that 
\begin{align} \label{BF21}
    BF \approx exp\left\{ -\frac{1}{2}(BIC_u - BIC_c) \right\}
\end{align} 
 to form the posterior model probability in (\ref{pcalib}). See \cite{KassRaftery1995,  Kass1995Reference} for more information about this approximation.  Here, the BIC under the well calibrated model $M_c$ is defined as 
 \begin{align} \label{BIC1}
BIC_c &= - 2 \times log(\pi(\delta = 1, \gamma =1|\mathbf{x},\mathbf{y})).
\end{align}
The penalty term for number of estimated parameters is omitted in (\ref{BIC1}) as both parameters are fixed at 1 under $M_c$. The BIC under poorly calibrated model $M_{nc}$ is defined as 
\begin{align} \label{BIC2}
BIC_u &= 2\times log(n) - 2\times log(\pi(\hat\delta_{MLE}, \hat\gamma_{MLE}|\mathbf{x},\mathbf{y})).
\end{align}
With this approximation for $BF$, we form $P(M_c|\mathbf{y})$, which can be interpreted as the probability the set of forecasts $\mathbf{x}$ is well calibrated given the observed data $\mathbf{y}$.  Again, $P(M_c|\mathbf{y})$ corresponds to calibration as $\delta=\gamma=1$ implies events happen at the rate forecasted with no further adjustment.  The interpretability of the posterior model probability, $P(M_c|\mathbf{y})$, is the key feature of this Bayesian test for calibration.  By quantifying the calibration of probability forecasts with a readily interpretable metric, we enable easier comparison of forecasters in terms of calibration and more informed decision making. We posit that $P(M_c|\mathbf{y})$ is interpretable to the extent that the Bayesian posterior probabilities that condition on observed data are interpretable.  For a frequentist approach to assessing calibration using (\ref{likelihood}), see the Likelihood Ratio test presented in the online supplement.

\subsection{Boldness-Recalibration}
\label{sec:br}

The previous Bayesian model selection approach assesses calibration alone with no regard for boldness.  We now consider the boldness of predicted probabilities measured by their spread (i.e. standard deviation)
\begin{align} \label{sb}
    s_b &= sd(\mathbf{x}).
\end{align}
The goal of boldness-recalibration is to maximize $s_b$, or boldness of predictions subject to a user-specified constraint on the calibration probability, $P(M_c|\mathbf{y})$.  To accomplish this we let the user set the calibration level, $t$, that $P(M_c|\mathbf{y})$ must achieve.  For example, if we want to ensure our recalibrated probabilities had a posterior probability of at least 95\%, we would set $t=0.95$.  Then boldness ($s_b$) is maximized subject to $P(M_c|\mathbf{y})=0.95$.  We call $\mathbf{x_t}$ the (100 * t)\% Boldness-Recalibration set where $x_{i,t} = c(x_i; \hat \delta_t, \hat \gamma_t)$ and 
\begin{align} \label{opt}
    (\hat \delta_t,\hat \gamma_t) &= argmax_{(\hat \delta,\hat \gamma)}(s_b:P(M_c|\mathbf{y}, \hat \delta,\hat \gamma)\geq t).
\end{align}

\vspace*{-0.3cm}
\begin{figure}[h!]
\begin{center}
\includegraphics[width=5in]{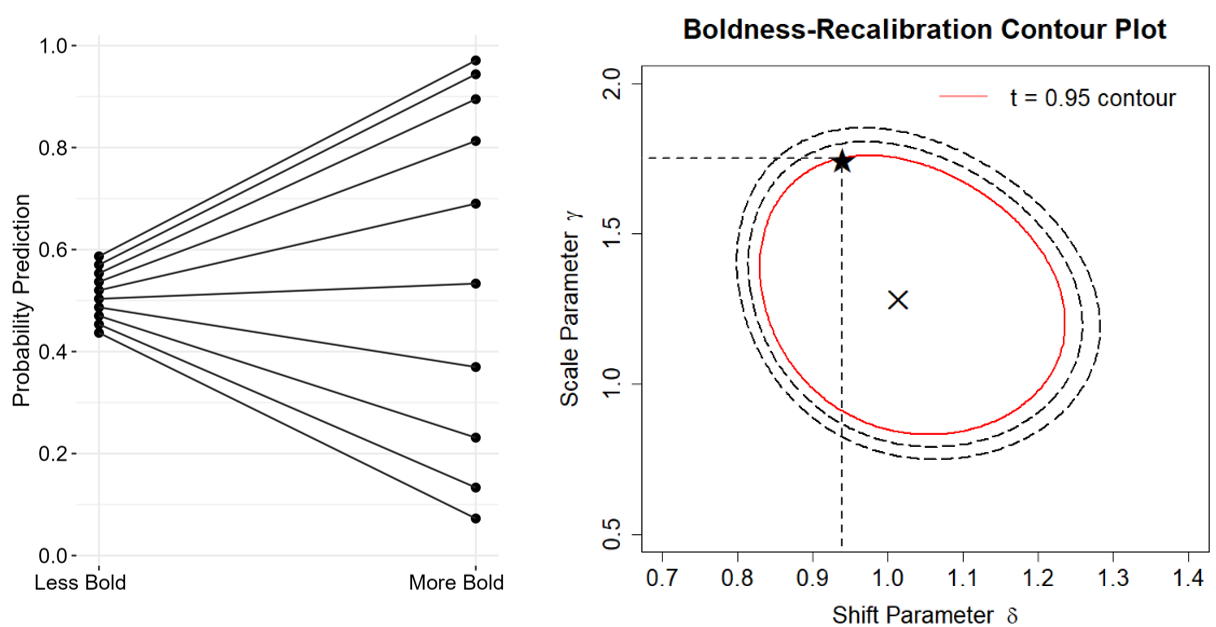}
\end{center}
\vspace*{-0.3cm}

\caption{Schemas to visualize boldness-recalibration.  The left panel shows boldness as a function of spread in predictions.  Each line corresponds to a prediction. The right panel shows a boldness-recalibration contour plot where the x-axis is shift parameter $\delta$, y-axis is scale parameter $\gamma$, and z-axis is $P(M_c|\mathbf{y})$ achieved by $\delta$ and $\gamma$.  Contours correspond to $P(M_c|\mathbf{y})$ = 0.95 (solid red), 0.9 and 0.8 (dashed black).  The $\times$ corresponds to ($\hat\delta_{MLE}$, $\hat\gamma_{MLE}$) such that the resulting probabilities under LLO-adjustment have maximal probability of calibration. The star on the 0.95 contour corresponds to ($\hat\delta_{0.95}$, $\hat\gamma_{0.95}$) such that the resulting probabilities have maximal spread subject to 95\% calibration.  These LLO-adjusted probabilities are called the 95\% boldness-recalibration set.  \label{fig:schemas}}
\end{figure}

To visualize the process of boldness-recalibration, consider the two schemas in Figure \ref{fig:schemas}.  The panel on the left depicts predictions that vary in boldness.  The ``less bold'' predictions are closer to the base rate $\bar y$.  The ``more bold'' predictions arise by moving the original predictions away from the base rate, and thus increasing spread.

The panel on the right of Figure \ref{fig:schemas} shows a boldness-recalibration contour plot.  This plot is used in the case study to show $P(M_c|\mathbf{y})$ across a grid of LLO-adjustment parameters $\delta$ and $\gamma$.   Rather than focus solely on where $P(M_c|\mathbf{y})$ is high (i.e. high calibration), we can draw a contour at $P(M_c|\mathbf{y}) = t$ to focus on our user specified level of calibration.  Then, along that contour we identify the $\delta$ and $\gamma$ that maximize spread in the LLO-adjusted probabilities via a grid-search based approach. The $\delta$ and $\gamma$ values corresponding to the star indicate precisely how to use Eq.(\ref{llo}) to embolden predictions subject to $t$. In Figure \ref{fig:schemas}, we identify these parameters with the star along the red contour at $t=0.95$. We call the LLO-adjusted probabilities under these parameters the 95\% boldness-recalibration set.

\subsection{Other Methods to Assess Calibration}
\label{sec:othermeth}

We report the Brier Score and Expected Calibration Error for the examples in this paper.  

\subsubsection{Brier Score Calibration Component}
\label{sec:bs}

For binary events, the Brier Score takes on the form 
\begin{align} \label{bs}
BS = \frac{1}{n} \sum_{i=1}^n (x_i - y_i)^2
\end{align}
where $x_i$ is the predicted probability for event $i$ and $y_i$ is the binary outcome (0 if a non-event, 1 if event).  The Brier Score in (\ref{bs}) can take on any value from 0 to 1, where lower values are better. 

The Brier Score can be decomposed as follows: 
\begin{align} \label{decomp}
BS = \frac{1}{n} \sum_{k=1}^K n_k (x_k  - \bar{y}_k)^2 - \frac{1}{n} \sum_{k=1}^K n_k (\bar{y}_k - \bar{y})^2 + \bar{y}(1-\bar{y})
\end{align} 
where $x_k$ is the average prediction for bin $k$, $\bar{y}_k$ is the relative frequency of events corresponding to the observations in bin $k$, $\bar{y}$ is the overall base rate, $K$ is the total number of bins, $n_k$ is the number of observations within bin $k$, and $n$ is the total number of predictions \citep{Murphy1973a}.  The first addend on the right hand side of (\ref{decomp}) is a measure of calibration, which we will refer to as Brier Score Calibration (BSC), and is the measure we will compare to $P(M_c|y)$.  The second addend on the right hand side of (\ref{decomp}) is a measure of resolution, which we will abbreviate as BSR, and is a measure we will compare to $s_b$.   The third addend is a measure of uncertainty in the outcomes, which we will abbreviate as BSU. Lower values of BSC are better, with BSC = 0 indicating perfect calibration.  Higher values of BSR are better, with BSR = BSU indicating perfect resolution.
 
\subsubsection{Expected Calibration Error (ECE)}

For binary events, Expected Calibration Error (ECE) takes on the form 
\begin{align} \label{ece}
ECE =  \sum_{k=1}^K \frac{n_k}{n} |\bar{y}_k - \bar{x}_k|
\end{align}
where $K$ is the number of bins, $n_k$ is the number of predictions partitioned into bin $k$, $\bar{y}_k$ is the proportion of observed events in bin $k$, and $\bar{x}_k$ is the average probability prediction in bin $k$.  ECE can take on any value from 0 to 1, where lower values are better.  

\subsection{Data}
\label{sec:data}

\subsubsection{Hockey Home Team Win Predictions}
\label{sec:hockeypundits}

To demonstrate the capabilities of boldness-recalibration, we assembled data from FiveThirtyEight that pertain to the 2020-21 National Hockey League (NHL) Season.  FiveThirtyEight produced predicted probabilities for all 868 regular season games that season via modelling with carefully constructed components based on expert knowledge of the game of hockey. These predictions were furnished prospectively pre-game, with no in- or post-game updating.   FiveThirtyEight probabilities are potentially hedged towards the base rate of 0.53 with an inter-quartile range of 0.12 (0.47, 0.59), their full range being (0.26, 0.77). More detailed information about this data set can be found in the online supplement. 

 In addition to this real-world forecaster, we generated a set of 868 random probability predictions to represent a hockey forecaster who is completely uninformed about the NHL games they aimed to predict. We call this forecaster our ``random noise forecaster." To mimic this behavior and better enable comparability, our random noise forecaster is generated by taking random uniform draws from 0.26 to 0.77, the observed range in the FiveThirtyEight data. The purpose of the random noise forecaster is to demonstrate how boldness recalibration operates when predictions are unrelated to the events they predict.  We want to ensure our method does not blindly embolden inaccurate forecasts.

\subsubsection{Simulation Study}

Table \ref{tab:settings} shows the four forecaster types in our simulation study.  The data generation process for our simulation follows:
\begin{enumerate}
    \item Generate $n$ true event outcomes via random independent Bernoulli draws, where the probability of success at each draw takes on a random uniform draw from 0 to 1.  $$p_i \sim \text{Uniform}(0,1)  \text{, } y_i \sim \text{Bernoulli}(p_i) \text{, } i=1,...,n$$
    The $p_i$s make up the well calibrated forecaster predictions by construction, as they directly correspond to the true probability of each event outcome.
    \item To manipulate classification accuracy, add varying amounts of random noise, $v_i$, to each $p_i$ on the log odds scale, which is equivalent to $$p_{i,\sigma} = \frac{e^{v_i}p_i}{(1-p_i)+e^{v_i}p_i}$$ where $p_{i,\sigma}$ is the set of noisy probabilities and $v_i \sim N(0, \sigma^2) \text{, } \sigma \in \{0 , 0.1, 0.5, 1, 2\}.$
    \item To manipulate boldness and create the four forecaster types, LLO-adjust $p_{i,\sigma}$ under varying $\delta$ and $\gamma$ values, summarized in Table \ref{tab:settings}.  Since the LLO function is monotone, forecasters LLO-adjusted from $p_{i,\sigma}$ maintain the same classification accuracy as $p_{i,\sigma}$.
\end{enumerate}

\begin{table}[h!]
    \centering
\begin{tabular}{ | m{9em} ||  m{1cm}  m{1cm} m{1cm}  m{1cm} m{11em} |}
\hline
 \textbf{Forecaster Type}  & $\mathbf{\delta}$ & $\mathbf{\gamma}$ & $\mathbf{\alpha}$ & $\mathbf{\beta}$ & \textbf{Description} \\
\hline\hline
 \textbf{Well Calibrated}  & 1  & 1 & 1  & 1  & Ideal Forecaster  \\ 
 \hline
 \textbf{Hedger }    & 1  & 0.25 & 0.25  & 0.76 & Not Bold \\ 
 \hline
 \textbf{Boaster}  & 1  & 2 & 2  & 1.44 & Too Bold \\
 \hline
 \textbf{Biased}  & 2 & 1 & 1  & 0.64 & Too Optimistic for Event \\ 
 \hline
\end{tabular}
    \vspace*{0.3cm}
    \caption{Values of LLO parameters ($\delta$, $\gamma$) and Prelec parameters ($\alpha$, $\beta$) under which each forecaster type is simulated along with description of forecaster type.}
    \label{tab:settings}
\end{table}

The first forecaster type, called Well Calibrated, represents forecasters whose predictions correspond to the true event rate.  Notice in Table \ref{tab:settings} that these predictions are LLO-adjusted under $\delta=\gamma=1$, so the Well Calibrated Probabilities are equivalent to the perfectly calibrated probabilities with added noise, i.e. $p_{i,\sigma}^{wc} = p_{i,\sigma}$.  Thus under $\sigma=0$, $p_{i,0}^{wc} = p_i$, the perfectly calibrated probabilities.  

Our second forecaster type is called Hedger. The Hedger compresses probabilities around the base rate, 0.5 in this case. We call their predictions ``hedged'' as they reflect forecasters who are lacking boldness even though their accuracy could be high. In contrast, our third forecaster type, Boaster, represents a forecaster who exhibits excessive boldness.  The majority of their predictions are far from the base rate and very close to the extremes of 0 and 1. The fourth forecaster type is Biased. These forecasters systematically make predictions that are higher or lower than the event rate. 

While this simulation focuses on miscalibration simulated via LLO-adjustment, we also explore miscalibration simulated from Prelec's two parameter function: 
\begin{equation}\label{prelec}
    w(\mathbf{x};\alpha, \beta) = e^{-\beta(-log(\mathbf{x}))^\alpha}
\end{equation}
where $\alpha > 0$ and $\beta > 0$ \citep{Prelec1998}.  We follow the same simulation procedure as shown above, except using (\ref{prelec}) rather than LLO in Step 3.  Similar to LLO-adjustment, the Well Calibrated forecaster is Prelec-adjusted under $\alpha = \beta = 1$.  The other forecaster parameter values for $\alpha$ and $\beta$ are summarized in Table \ref{tab:settings}.  Note the original formulation of this function in \cite{Prelec1998} limits $0 < \alpha < 1$, we allow $\alpha \geq 1$ as the function provides valid probabilities under these settings. Given that our methodology assumes miscalibration follows the LLO function which likely will not hold in all scenarios, we simulate miscalibration via the Prelec function to assess how well our methodology does until miscalibration misspecification. 

To explore the effect of sample size on our methodology, we generated data sets of size $n=$ 30, 100, 800, 2,000, and 5,000.  In total, we present results from 100 Monte Carlo (MC) replicates for each value of $n$.  Throughout the study, one MC replicate consists of a set of $n$ outcomes, and a corresponding set of $n$ predicted probabilities for each of the four forecasters types under each of the five noise settings and both  LLO and Prelec adjustment (35 total predicted probabilities sets for each replicate).  For each of the 35 sets, we (i) LLO-adjust via MLEs $\hat\delta_{MLE}$, $\hat\gamma_{MLE}$, (ii) 95\%, 90\%, and 80\% boldness-recalibrate, and then (iii) evaluate the calibration and boldness of the adjusted sets from (i) and (ii).

\section{Results}
\label{sec:res}

\subsection{Hockey Home Team Win Predictions Case Study}
\label{sec:hockeyres}

\begin{figure}[h!]
\begin{center}
\vspace*{-0.3cm}
\includegraphics[width=6.25in]{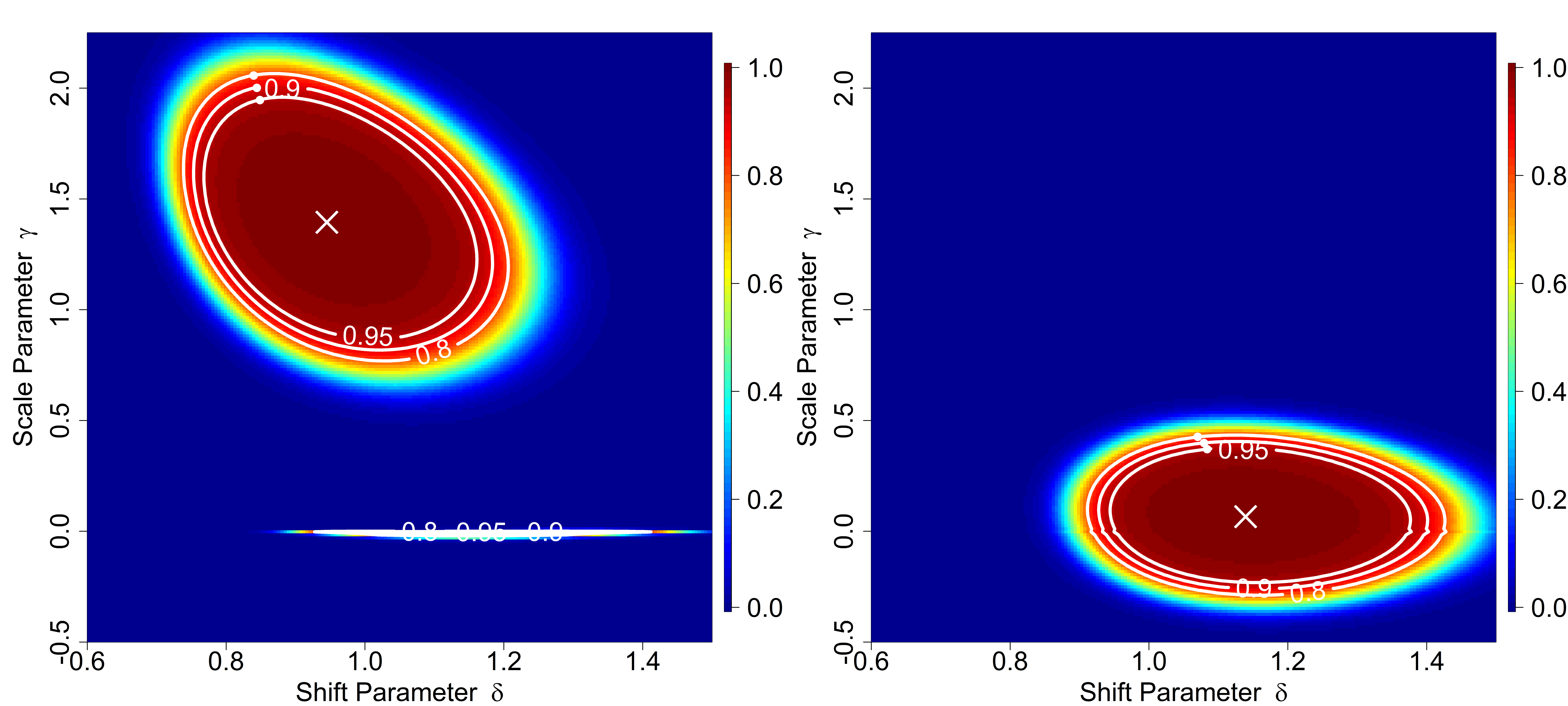}
\end{center}
\vspace*{-0.3cm}
\caption{ Boldness-recalibration contour plots for FiveThirtyEight (Left) and random noise forecaster (Right). Regions in red reflect high $P(M_c|\mathbf{y})$ for the LLO-adjusted $\mathbf{x}$ via corresponding $\delta$ (x-axis) and $\gamma$ (y-axis) values.  Regions in blue show low $P(M_c|\mathbf{y})$. The $\times$ marks $\hat\delta_{MLE}$ and $\hat\gamma_{MLE}$ where the probability of calibration is maximized.  Contours at $t=0.95, 0.9,$ and $0.8$ are drawn in white and $\hat\delta_{t}$ and $\hat\gamma_{t}$ are marked by white points along each contour. \label{fig:hockey_contours}}
\end{figure}

We applied boldness-recalibration to the two Hockey forecasters at three specified levels of calibration: $t=0.95, 0.9,$ and $0.8$.  Figure \ref{fig:hockey_contours} shows the boldness-recalibration plots for FiveThirtyEight (Left) and random noise forecaster (Right).  Regions in red show where $P(M_c|\mathbf{y})$ is high for the LLO-adjusted $\mathbf{x}$ via the corresponding $\delta$ (x-axis) and $\gamma$ (y-axis) values.  Regions in blue show where $P(M_c|\mathbf{y})$ is low.  As expected, $\hat\delta_{MLE}$ and $\hat\gamma_{MLE}$, marked by the white $\times$ in Figure \ref{fig:hockey_contours}, lie at the point where the probability of calibration is maximized.  The values for $\hat\delta_{t}$ and $\hat\gamma_{t}$ are marked by white points along the contour for each $t$.  Recall these represent the set of LLO-adjustment parameters for which maximal boldness is achieved with a probability of calibration of at least $t$.  These parameter values, along with the achieved $P(M_c|\mathbf{y})$, $s_b$, prediction range, Brier Score, BSC, BSR, ECE, and AUC are summarized in Table \ref{tab:hockeyparams}.

\textbf{\begin{table}[h!]
    \centering
    \small
    \settowidth\rotheadsize{\textbf{FiveThirtyEight}}
    \begin{tabular}{c @{\hspace{0.75ex}} l | r r r @{\hspace{0.1cm}} r r r r r r r r}
    \hline
     & & $P(M_c|\mathbf{y})$   & $s_b$ & \multicolumn{2}{>{\centering\arraybackslash\setlength{\baselineskip}{0.5\baselineskip}}b{2cm}}{Prediction Range}  & BS & BSC  & BSR & ECE & AUC  & $\mathbf{\hat\delta}$ & $\mathbf{\hat\gamma}$ \\ 
     \hline
     \multirow{5}{*}{\rothead{\textbf{FiveThirtyEight}}}
     & \textbf{Original} &  0.9904 & 0.091 & 0.26 & 0.78 & 0.236 &  0.002 & 0.015 & 0.052 & 0.65 & - & - \\
     & \textbf{MLE} &  0.9988 & 0.124  & 0.18 & 0.84 &  0.235 & 0.001 & 0.014 & 0.038 & 0.65 &  0.95 & 1.40\\
     & \textbf{95\% B-R}  &  0.9500 & 0.165  &  0.10 &  0.91 & 0.233 & 0.002 & 0.018  & 0.055 & 0.65 & 0.87 & 1.96\\
     & \textbf{90\% B-R}  &  0.9000 & 0.169 & 0.10 & 0.91 & 0.233 & 0.002 & 0.018& 0.057 & 0.65 &  0.87 & 2.01\\
     & \textbf{80\% B-R}  &  0.8000 & 0.173 & 0.09 & 0.92 & 0.234 & 0.003 & 0.018 & 0.058 & 0.65 & 0.86 & 	2.07 \\
     \hline 
     \multirow{5}{*}{\rothead{\textbf{Random Noise}}}
     & \textbf{Original} &  0.0000 & 0.146 & 0.26 & 0.77 & 0.267 &  0.019 & 0.001 & 0.124 & 0.51 & - & - \\
     & \textbf{MLE} &  0.9988 & 0.011  & 0.51 & 0.56 &  0.249 & 0.000 & 0.000 & 0.021 & 0.51 &  1.14 & 0.07\\
     & \textbf{95\% B-R}  &  0.9500 & 0.058  &  0.43 & 0.64 & 0.251 & 0.002 & 0.000  & 0.049 & 0.51 & 1.12 & 0.38\\
     & \textbf{90\% B-R}  &  0.9000 & 0.062 & 0.42 & 0.65 & 0.251 & 0.002 & 0.000 & 0.052 & 0.51 &  1.12 & 0.41\\
     & \textbf{80\% B-R}  &  0.8000 &  0.066 & 0.42 & 0.66 & 0.252 & 0.003 & 0.000 & 0.056 & 0.51 & 1.12 & 0.44\\
     \hline 
    \end{tabular}
    \vspace*{0.3cm}
    \caption{Values of the posterior model probability of calibration $P(M_c|\mathbf{y})$, boldness ($s_b$), prediction range, Brier Score (BS), Brier Score calibration component (BSC), Brier Score resolution component (BSR), expected calibration error (ECE), and area under the ROC curve (AUC) for the original sets of predictions and those achieved under MLE recalibration, 95\%, 90\%, and 80\% boldness-recalibration (B-R) via estimated adjustment parameters $\hat \delta$ and $\hat \gamma$ for FiveThirtyEight and random noise forecaster.}
    \label{tab:hockeyparams}
\end{table}}

\begin{figure}[h!]
\begin{center}
\includegraphics[width=6in]{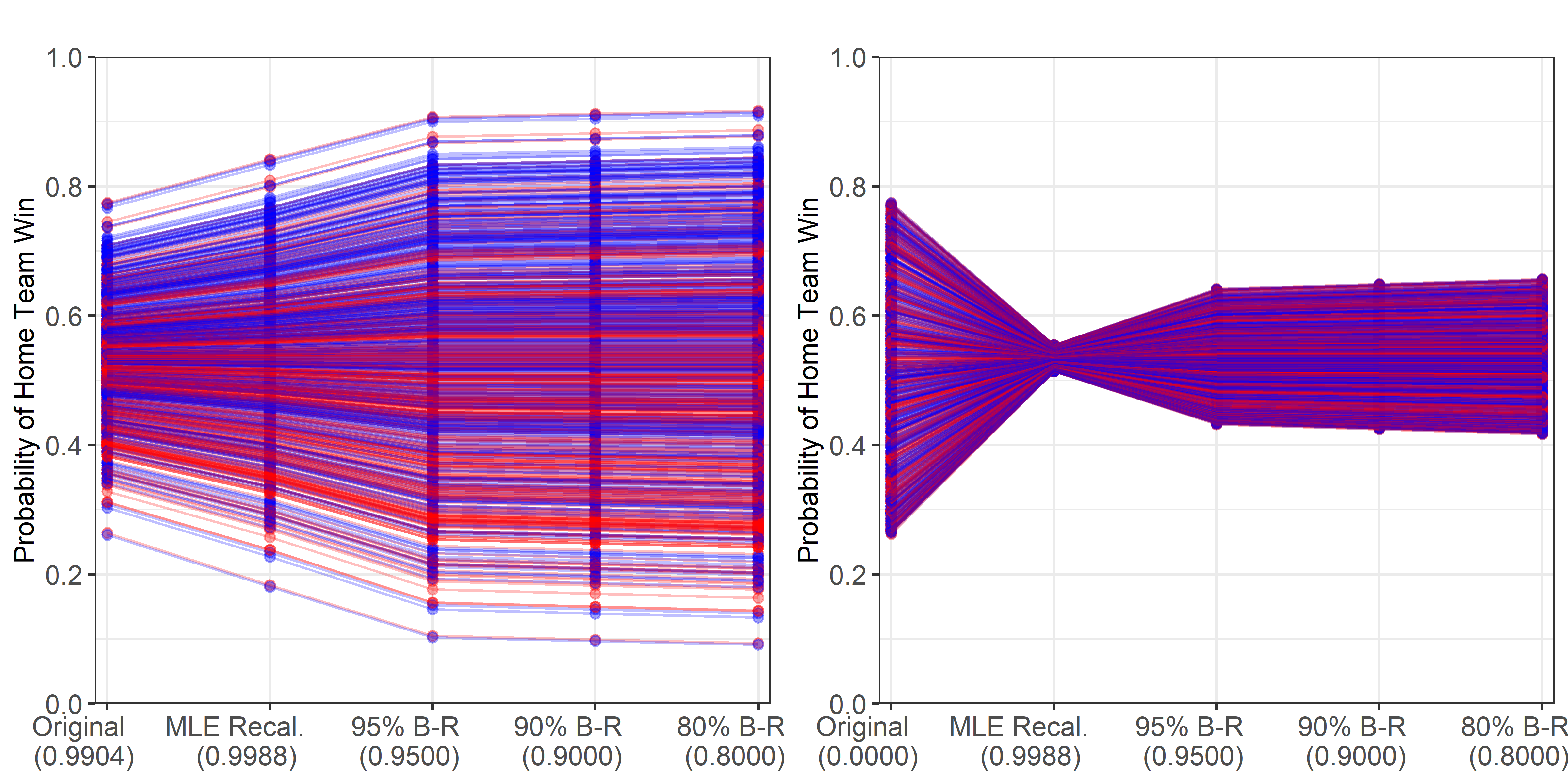}
\vspace{-0.5cm}
\end{center}
\caption{ Lineplot visualizing how the predictions for FiveThirtyEight (Left) and the random noise forecaster (Right) change under LLO-adjustments via MLEs and boldness recalibration. The first column of points in each panel is the original set of probability predictions.  The second column of points is the predictions after recalibrating with the MLEs.  The last three columns are the predictions after 95\%, 90\%, and 80\% boldness-recalibration respectively. A line is used to connect each original prediction to where it ends up after each recalibration procedure. Points and lines colored blue correspond to predictions for games in which the home team won.  Red corresponds to games in which the home team lost. Achieved $P(M_c|\mathbf{y})$ is reported in the parentheses in the axis label. \label{fig:hockey_lines}}
\end{figure}

After deploying boldness-recalibration on the two sets of predictions, we see a substantial increase in boldness for FiveThirtyEight.  Figure \ref{fig:hockey_lines} shows how the predictions for FiveThirtyEight (Left) and the random noise forecaster (Right) change under LLO-adjustments via MLEs and boldness recalibration. The first column of points in each panel represents the original set of probability predictions.  The second column of points represents the predictions after recalibrating with the MLEs.  The third, fourth, and fifth columns of points represent the predictions after 95\%, 90\%, and 80\% boldness-recalibration respectively. A line is used to connect each original prediction to where it ends up after each recalibration procedure. Points and lines colored blue correspond to predictions for games in which the home team won.  Red corresponds to games in which the home team lost. The posterior model probability of calibration is reported in the parentheses in the axis label.  Note that the posterior model probabilities are not necessarily linear from left to right.  We order the sets in this way, starting with the original forecasters sets, to make consistent comparisons throughout the results of the paper.

First, notice that the probability of calibration given the event outcomes for the original FiveThirtyEight forecasts is very high at 0.9904, whereas the probability for the random noise forecaster rounds down to 0.000.  This indicates that FiveThirtyEight is well calibrated to begin with and, as we would expect, the random noise forecaster is not.  Next, notice that by maximizing $P(M_c|\mathbf{y})$, the range of FiveThirtyEight's predictions expands from (0.26, 0.77) to (0.18, 0.84) and $s_b$ increases from 0.091 to 0.124 as seen in Table \ref{tab:hockeyparams}.  FiveThirtyEight can achieve a maximal probability of calibration of 0.9988. In contrast, for the random noise forecaster to achieve their maximal calibration of 0.9988, they must pull their predictions in toward the base rate of 0.53.  Their prediction range contracts from  (0.26, 0.77) to (0.51, 0.56) and $s_b$ drops from 0.146 to 0.011. Not only does this imply the random noise forecaster is poorly calibrated, but it also suggests that their predictions do not have useful predictive information.  We know this to be true because these predictions were randomly generated with no association with the outcome. 

Now compare the spread of original predictions to the spread of the 95\% boldness-recalibration set.  FiveThirtyEight can further embolden their predictions by accepting a 5\% risk of mis-calibration, expanding their range to (0.10, 0.90).  This suggests that FiveThirtyEight could embolden predictions with a modest decrease in $P(M_c|\mathbf{y})$, where the random noise forecaster has no knowledge of the outcome and should make far more cautious calls.   In this example, there is minimal gain in boldness moving from 95\% B-R to 90\% or 80\%.  It is up to the discretion of the user to determine if accepting an additional risk of 5\% or 10\% risk of miscalibration is worth the minimal reduction in boldness.  Regardless, boldness-recalibration successfully increases the boldness of our skilled hockey forecaster while maintaining a user-specified level of calibration.  For our random noise forecaster, boldness-recalibration suggests that increasing boldness would not be responsible, and instead contracts predictions.

In terms of the Brier Score, FiveThirtyEight and the random noise forecaster achieve scores of 0.2346 and 0.2675 respectively.  It is hard to say how this practically translates to how much ``better'' FiveThirtyEight is compared to the random noise forecaster and what a ``good'' Brier Score is for this application.   Despite the substantial increase in $s_b$ and prediction range for FiveThirtyEight, the Brier Score shows very little change, improving by 0.001 under MLE recalibration and an additional 0.002 under 95\% B-R. The BSC is the same for the original and 95\% B-R sets and BSR improves by 0.003.  In contrast, BSC for the random noise forecaster drops to near zero after MLE recalibration and then worsens to 0.002 under 95\% B-R. The BSR worsens after MLE recalibration and remains at  0.000, the worst achievable score for BSR, for all B-R sets. This further reflects that the random noise forecaster can only improve calibration by reducing boldness and resolution.  We see that ECE is minimized when calibration is maximized for both forecasters and worsens by 0.017 under 95\% B-R for FiveThirtyEight and by 0.028 for the random noise forecaster.

In terms of classification accuracy, FiveThirtyEight produces an AUC of 0.65.  This implies their predictions are better than chance and provide some information in classifying a home team win.  Our random noise forecaster produces an AUC of 0.51, which is very close to the underlying 0.5 we would expect as this forecaster makes predictions completely via random chance. Notice that AUC stays the same across all sets for both forecasters.  This is due to the fact that the LLO function is a monotonic function and the ordering of predictions does not change and neither does AUC.

\subsection{Simulation Study}
\label{sec:simres}

\begin{figure}[h!]
\begin{center}
\includegraphics[width=6in]{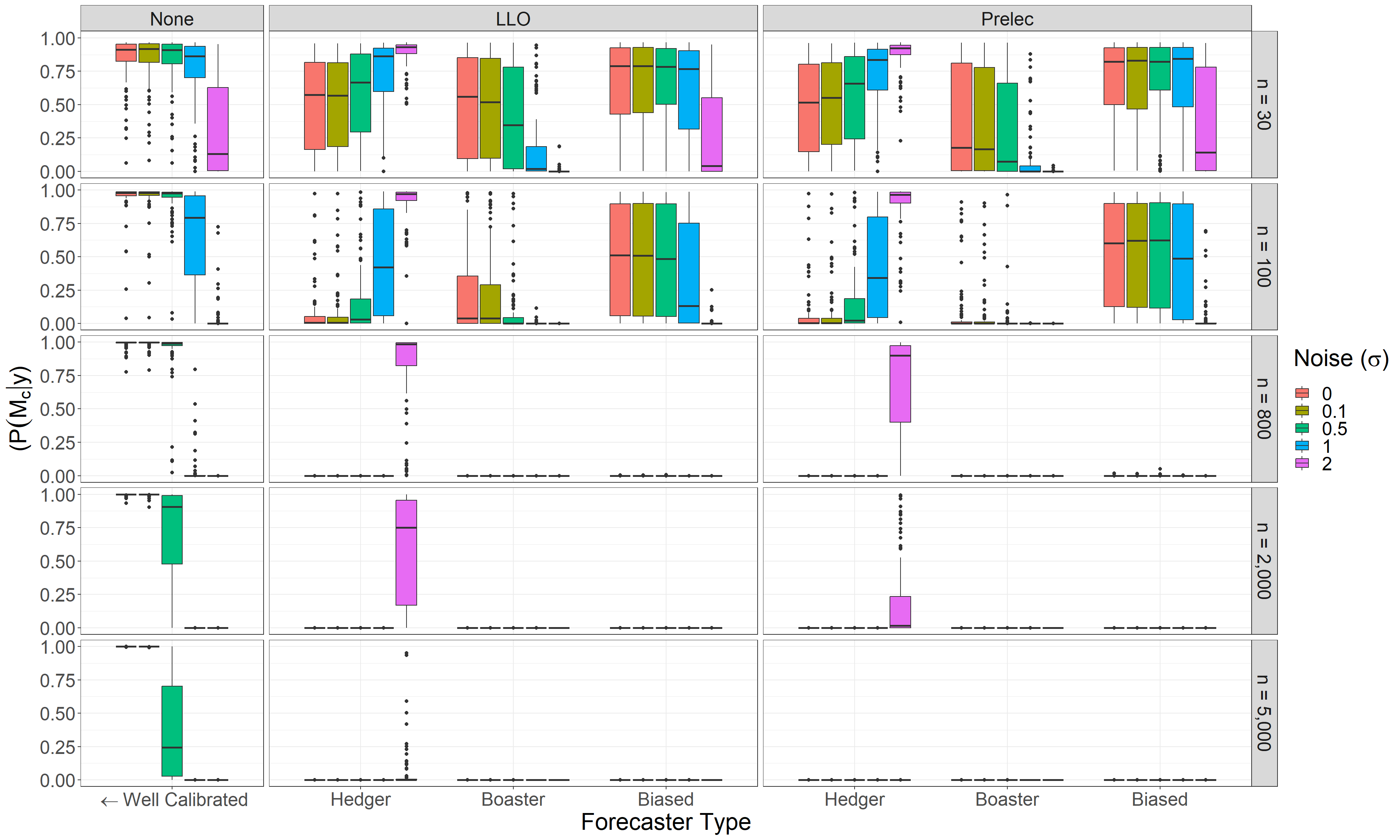}
\vspace{-0.5cm}
\end{center}
\caption{Boxplots summarizing the posterior model probability of calibration, $P(M_c|\mathbf{y})$, on the y-axis for 100 MC runs on simulated forecasters.   Boxplots are grouped by forecaster type on the x-axis.  Within groups, added noise increases from left to right.  Only the leftmost boxplot in the Well Calibrated group is perfectly calibrated, and ← indicates calibration increases as noise decreases.  Horizontal panels indicate which adjustment (if any) was applied to create the forecaster type.} As sample size increases with vertical panels, $P(M_c|\mathbf{y})$ decreases for all forecasters except Well Calibrated with little to no added noise.   \label{fig:pmp_box}
\end{figure}

The posterior model probability of calibration, $P(M_c|\mathbf{y})$, prior to MLE recalibration and boldness-recalibration is summarized in Figure \ref{fig:pmp_box} for all 100 MC runs. The boxplots are grouped by simulated forecaster type shown on the x-axis.  The y-axis shows the value of $P(M_c|\mathbf{y})$.  Sample size increases with vertical panels from top to bottom. Horizontal panels indicate whether miscalibration was simulated under the LLO or the Prelec function.  As the Well Calibrated forecasters do not change under either LLO or Prelec adjustment, they are separated out into their own panel. Within each group of boxplots, added noise $\sigma$ increases from left to right.  Thus, for the Well Calibrated group, only the first boxplot with no added noise ($\sigma=0$) is perfectly calibrated and both calibration and accuracy decrease as noise increases. Boxplots for $s_b$, BS, BSR, BSC, and ECE can be found in the online supplement. 

We expect $P(M_c|\mathbf{y})$ to be high for well calibrated forecasters and low for poorly calibrated forecasters.  As sample size increase, $P(M_c|\mathbf{y})$ decreases for all settings except the Well Calibrated forecasters with little to no added noise.  This indicates our Bayesian approach performs sensibly in that the ability to correctly detect miscalibration increases with sample size.  Additionally, as more noise is added to the Well Calibrated forecaster, their probability of calibration decreases as expected.  Notice that under low sample sizes, Hedgers with large added noise appear well calibrated.  This is not surprising as we have already established that hedging predictions with poor classification accuracy is a favorable strategy to achieve calibration. Also notice that under miscalibration misspecification (i.e. we assume miscalibration follows LLO when it actually follows Prelec), we see similar, if not improved, detection of miscalibration.

All simulated prediction sets were MLE recalibrated, and 95\%, 90\%, and 80\% boldness-calibrated.  Of the 17,500 total prediction sets,  95\%, 90\%, and 80\%  boldness-recalibration was successful in 99.4\%, 99.2\%, and 98.7\% of cases, respectively. By ``successful'', we mean that these sets were maximally emboldened while calibration of t = 0.95, 0.9, or 0.8 was maintained.  In most of the small percentage of cases where boldness-recalibration was not successful, the underlying optimization was unable to converge to parameters that achieved the desired level of calibration.  In the other few cases, adding random noise to the probabilities caused perfect separation of events and non-events.  Under MLE recalibration, these predictions are all moved to either 0 or 1, where no further emboldening is possible. The sets where boldness-recalibration was not achievable were removed from the results, as our focus is demonstrating the capabilities of boldness-recalibration.  

\begin{figure}[h!]
\begin{center}
\includegraphics[width=6.5in]{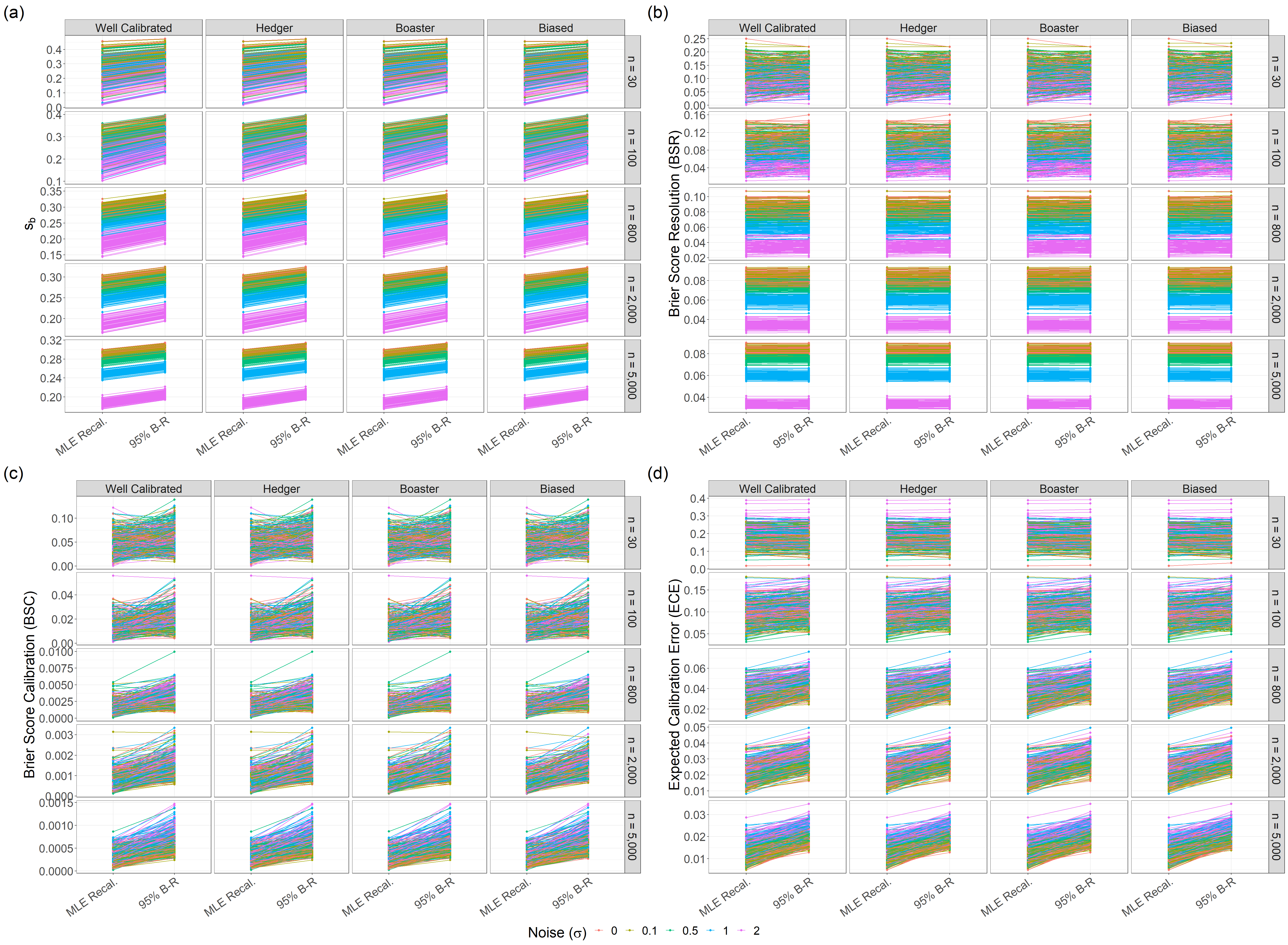}
\vspace{-0.5cm}
\end{center}
\caption{ Lineplots summarizing on the y-axis the change in (a) boldness measured by $s_b$, (b) Brier Score Resolution, (c) Brier Score Calibration, and (d) Expected Calibration Error for 100 MC runs on LLO-miscalibrated simulated forecasters.  Sample size increase with vertical panel.  Horizontal panels denote the forecaster type.  The first column of points in each panel represents the value of each metric for the MLE recalibration set.  The second column of points represents the same metric for the 95\% B-R set.  A line is used to connect points that correspond to the same original set of predictions.  The lines and points are colored based on the amount of added noise. Note that y-axis is not fixed across vertical panels, the points/lines are plotted in a randomized order, and one set is removed from each panel due to random perfect separation issues as described in the text.}   \label{fig:lp_comb}
\end{figure}

Figure \ref{fig:lp_comb} summarizes the change in $s_b$, BSR, BSC, and ECE moving from the MLE recalibrated set to the 95\% B-R set under LLO miscalibration.  These lineplots are different from the lineplots from the Hockey example in that the y-axis shows the value of the metric of interest.  Sample size increase with vertical panel.  Horizontal panels denote the forecaster type.  It is important to note the y-axis is not fixed across vertical panels.  The first column of points in each panel represents the value of each metric for the MLE recalibration set.  The second column of points represents the same metric for the 95\% B-R set.  A line is used to connect points that correspond to the same original set of predictions.  The lines and points are colored based on the amount of added noise.   We choose to only show the metric values under MLE recalibration and 95\% boldness-recalibration for two reasons: (i) it is best practice to not operate under poor calibration, whether it be the original set or sets at low boldness-recalibration thresholds like 80\% and (ii) we see little change in these metrics moving from the 95\% to the 90\% B-R set.  Additionally, we found that the results for the Prelec function were nearly identical to those for LLO miscalibration, so we focus on LLO here. Results for all sets can be found in the online supplement.

Notice in Figure \ref{fig:lp_comb}a that when moving from maximal calibration to 95\% calibration, all sets are emboldened to some degree.  However, as added noise increases, boldness decreases.  This is desirable as less noisy predictions should be more bold than more noisy predictions.  The distinction in $s_b$ between levels of added noise becomes more clear with higher sample sizes.  While the increase in $s_b$ from the MLE recalibrated set and 95\% B-R set diminishes with sample size, we expect that the degree to which emboldening is appropriate also diminishes.  Where data is abundant and proved to be extremely reliable, there may be less need for emboldening, as the predictions are already useful for decision making.

As for the Brier Score Resolution, notice in Figure \ref{fig:lp_comb}b that there is little change between MLE recalibration and 95\% B-R under small sample sizes.  In just a few cases under n = 30, we see that BSR decreases after emboldening.  There is virtually no change in BSR under large sample sizes.  As expected, with more added noise, BSR is typically lower.

Brier Score Calibration is much less consistent across MC runs than the other metrics. Notice in Figure \ref{fig:lp_comb}c that there is little to no distinction in BSC between levels of added noise.  Additionally, at small samples sizes BSC sometimes improves but other times worsens.  Under large sample sizes, BSC always worsens as we would expected given that we are sacrificing calibration in favor of boldness.  Similarly, in Figure \ref{fig:lp_comb}d we see that there is little distinction in expected calibration error between levels of added noise.  However, we see more consistency in terms of the degree of increase in ECE across MC runs.  Under small samples sizes, we see little to no change in ECE.  As sample size increases, we see larger increases in ECE moving from MLE recalibration to 95\% B-R.

\section{Conclusion}
\label{sec:conc}

This paper develops boldness-recalibration methodology surrounding the fundamental tension between calibration and boldness of predicted probabilities, subject to classification accuracy. While some methods consider concepts similar to boldness, such as resolution, relative to calibration, none provide direct control of the trade-off between the two. Our proposed Bayesian calibration assessment and boldness-recalibration approaches address this gap. This paper is for those who would consider making a small sacrifice in posterior probability of calibration to gain boldness so as to study riskier predictions for decision making.

The backbone of these approaches is the interpretable (in a Bayesian sense) posterior model probability, $P(M_c|\mathbf{y})$, which serves as a measure of calibration and is interpreted as the probability a set of predictions is calibrated, given the data observed. We define boldness as the spread (i.e. standard deviation) in predictions. In boldness-recalibration, the user pre-specifies a tolerable risk of miscalibration (e.g. $P(M_c|\mathbf{y})$ = 0.95) and subject to this constraint, our method maximizes spread in predictions and thus, boldness. The difference in the posterior model probabilities for the original and boldness-recalibrated sets concisely quantifies the calibration-boldness trade-off.  By pre-specifying calibration via $P(M_c|\mathbf{y})$, the user is given direct control of the boldness calibration trade-off.

Boldness-recalibration provides a means of appropriately emboldening probability predictions.  The Hockey case study shows that ``appropriate'' may have different meaning depending on the quality of the data.  The predictions from FiveThirtyEight were substantially emboldened while maintaining reasonable calibration.  This indicates their predictions are reliable but overly cautious. However, boldness-recalibration showed that the random noise forecaster should bring their predictions in towards the base rate rather than embolden. In this case, it is more appropriate to un-embolden as their predictions were not reliable or useful for decision making.

We demonstrate via simulation study that $P(M_c|\mathbf{y})$ correctly identifies miscalibration and appropriately emboldens in nearly all forecaster types at a reasonable sample size, even under miscalibration mispecification.  After correcting miscalibration in the simulated predictions, we see an increase in boldness across all sets.  In cases where the original predictions are noisy, spread is lower in the boldness-recalibrated set than in the original and higher in for those that are accurate. 

While we leverage spread to measure boldness and the LLO function to recalibrate, one could consider alternatives to these choices.  The core idea of selecting an emboldening plan that satisfies a required probability of calibration still holds.  Another potential future research goal of interest is that of subjective elicitation of prior probabilities of calibration.  While we use $P(M_c)$ = $P(M_u) = \frac{1}{2}$, this may not be ideal in situations where prior information can be obtained.  Additionally, another goal is to investigate potential dependence structure among forecasts, as this methodology does not currently enable analysis of dependence between forecasts in a set.  While we provide one example of a use-case in this paper, we propose these methods are useful in many situations where there are predicted probabilities of binary event. These methods allow decision makers to rely on these predictions for make informed decisions. Appropriately emboldened predictions, as produced by Boldness-recalibration in an interpretable manner, enable better decision making in these critical situations.\\

\noindent
{\textbf{Acknowledgements}}

\if0\blind
{
The authors would like to thank Chris Wilson, Damon Kuehl, Matthew Keefe, Andrew McCoy, Tyler Cody, Xin Xing, and Bill Woodall for their insights and roles obtaining case study data for this line of research. The authors are grateful for the helpful comments from the associate editor and two referees.
} \fi

\if1\blind
{
BLINDED
} \fi


\bigskip


\bibliographystyle{agsm}

\bibliography{Bibliography-PPC}

\noindent
\section*{Supplemental material for \textit{Boldness-Recalibration for Binary Event Predictions}}

\subsection{Likelihood Ratio Test for Calibration}
\label{sec:LRT}

 Let $\mathbf{y}$ be a vector of $n$ outcomes corresponding to the $n$ predictions in $\mathbf{x}$ from a single forecaster.  Then, we have
\begin{equation}\label{likelihood}
    \pi(\mathbf{x}, \mathbf{y} | \delta, \gamma) = \prod_{i=1}^n c(x_i;\delta, \gamma)^{y_i} \left[1-c(x_i;\delta, \gamma)\right]^{1-y_i}.
\end{equation}
Perhaps surprisingly, it does not appear as if a Likelihood Ratio Test (LRT) based on (\ref{likelihood}) has been previously proposed in the literature. We devise a frequentist approach to this problem by setting up a likelihood ratio test for the following hypotheses:
\begin{align*}
&H_0: \delta = 1, \gamma = 1 \text{   (Probabilities are well calibrated)}\\
&H_1: \delta \neq 1 \text{ and/or } \gamma \neq 1 \text{   (Probabilities are poorly calibrated)}
\end{align*}
Thus, the likelihood ratio test statistic for $H_0$ is
\begin{align} \label{TS}
\lambda_{LR} &= -2 log\left[\frac{\pi(\delta =1, \gamma=1|\mathbf{x},\mathbf{y})}{\pi(\delta = \hat\delta_{MLE}, \gamma = \hat\gamma_{MLE}|\mathbf{x},\mathbf{y})}\right],
\end{align}
where $\lambda_{LR}\stackrel{H_0}{\sim}{\chi^2_2}$ asymptotically under the null hypothesis $H_0$.  The above test statistic can be used to test the calibration of a vector of predicted probabilities, $\mathbf{x}$, and the vector of corresponding outcomes, $\mathbf{y}$.

\subsection{Case Studies}

\subsubsection{Hockey Home Team Win Predictions}
\label{sec:hockeydat}

FiveThirtyEight provided their predictions in the form of a downloadable spreadsheet with the predicted probabilities of a home team win or away team win on their website (\url{https://data.fivethirtyeight.com/}). Details on how FiveThirtyEight produced these probabilities can also be found on their website (\url{https://fivethirtyeight.com/methodology/how-our-nhl-predictions-work/}). The win/loss game results were obtained by web-scraping from NHL.com using the NHL API. These results were then mapped to the probabilities forecasts of FiveThirtyEight by matching by game date and teams.  FiveThirtyEight's predictions are compiled from year to year, so the data for the 2020-21 season is included in the same downloadable spreadsheet as the most recent season's predictions. This data set was used because it is publicly available online, it provided a complete seasons worth of game observations, and it display properties that we were particularly interested in exploring via our method.  

An important thing to note about the 2020-21 NHL season that distinguishes it from other seasons is that it comes on the heels of an abbreviated 2019-20 season due to the COVID-19 pandemic.  With the previous season significantly shortened and playoff pushed to the end of summer, the 2020-21 season started late in comparison to a ``normal" NHL season and consisted of only 56 games per team instead of the ``normal" 82 games per team.  Additionally, teams were restricted to only playing within their division, with divisions re-aligned to minimize travel.  While these abnormalities do not directly affect the validity of our methodology, they should be kept in mind when comparing forecaster predictions from this season to other seasons.

\subsection{Connection to Logistic Regression}
\label{sec:logreg}

Our approach of adopting a Bernoulli Likelihood governed by LLO-adjusted probabilities is equivalent to a specialized logistic regression model.  As established, our outcome $y \sim \text{Bernoulli}(p)$.  We have a linear predictor in the form of $\eta = X\beta = \gamma log\left( \frac{x}{1-x}\right) + \tau$. Lastly, we have the logit link such that $\eta = log\left(\frac{p}{1-p}\right)$ The key difference between standard logistic regression and our approach is that we let $p = c(x)$, or in words, we let the governing probabilities of our outcomes equal the LLO-adjusted probabilities from the LLO function.  

\subsection{Additional Simulation Study Results}

\begin{figure}[h!]
\begin{center}
\includegraphics[width=6.3in]{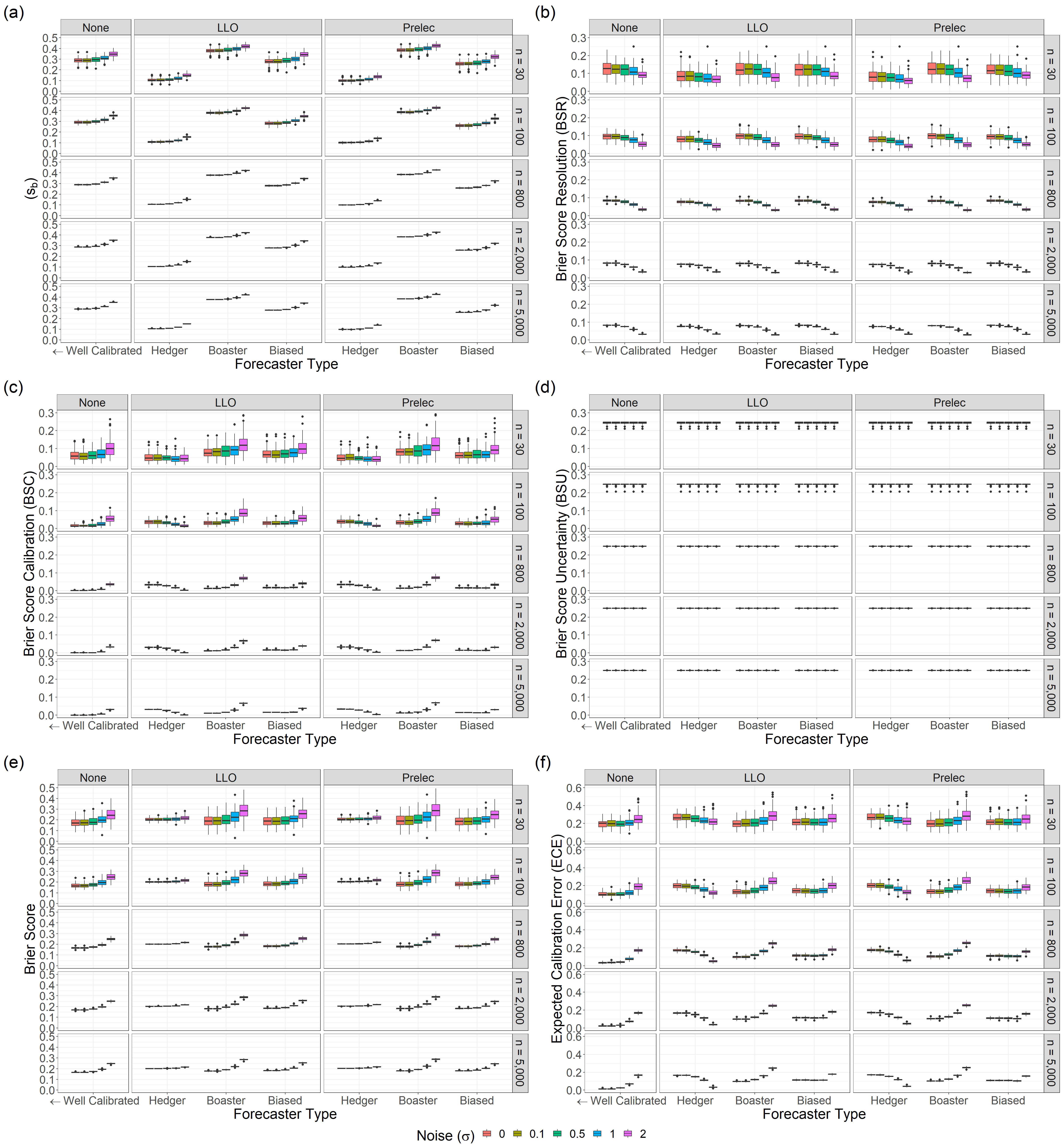}
\vspace{-0.5cm}
\end{center}
\caption{Boxplots summarizing (a) $s_b$, (b) the Brier Score Resolution, (c) Calibration, and (d) Uncertainty components, (e) overall Brier Score, and (f) Expected Calibration Error (y-axis) for 100 MC runs on simulated forecasters.   Boxplots are grouped by forecaster type on the x-axis.  Within groups, added noise increases from left to right.  Only the leftmost boxplot in the Well Calibrated group is perfectly calibrated, and ← indicates calibration increases as noise decreases.  Horizontal panels indicate which adjustment (if any) was applied to create the forecaster type. Sample size increases with vertical panels.  
\label{fig:comb_box}}
\end{figure}

Figure \ref{fig:comb_box} summarizes (a) $s_b$, the Brier Score  (b) Resolution, (c) Calibration, and (d) Uncertainty components, (e) overall Brier Score, (f) Expected Calibration Error  for each simulation setting prior to MLE recalibration and boldness-recalibration. The boxplots are grouped by simulated forecaster type shown on the x-axis.  The y-axis shows the value of the metric of interest.  Sample size increases with vertical panels from top to bottom. Within each group of boxplots, added noise $\sigma$ increases from left to right.  Thus, for the Well Calibrated group, only the first boxplot with no added noise ($\sigma=0$) is perfectly calibrated and calibration decreases as added noise increases.  Notice that there is little to no difference between the boxplots under LLO miscalibration compared to Prelec miscalibration.

In Figure \ref{fig:comb_box}a, as expected we see that $s_b$ is higher for Boasters and lower for Hedgers compared to the Well Calibrated forecasters.  We also see that $s_b$ is the same for Biased compared to Well Calibrated as the Biased probabilities are only shifted, not scaled.  Additionally, we see that the mean $s_b$ across MC runs does not change with sample size. However, in Figure \ref{fig:comb_box}b we see that the mean Brier Score Resolution Component (BSR) degrades (decreases) with sample size.  Additionally, the difference in BSR between the Well Calibrated Forecasters and the Hedgers and Boasters is not as clear as it is with $s_b$.

Figures \ref{fig:comb_box}c and \ref{fig:comb_box}f show that as sample size increases the Brier Score Calibration component (BSC) and Expected Calibration Error (ECE) improve (decrease) for all settings, not just those that are well calibrated.  At large sample sizes, the Hedgers with large added noise score nearly indistinguishable from the perfectly calibrated forecaster on the far left under BSC.  While BSC for the Boasters and Biased forecasters are higher than the well calibrated, the difference in minimal. ECE provides greater distinction between forecasters than BSC.

Notice that the Brier Score Uncertainty component (BSU) does not change across forecaster types (Figure \ref{fig:comb_box}d).  This is due to BSU not relying on the probabilitiy predictions themselves. Rather, it is simply a measure of the uncertainty in the outcomes.  Since BSU is a metric that cannot be changed by improving (degrading) forecasts and thus is out of the control of forecasters, we choose not to discuss BSU further.

Often it is more common to use the overall Brier Score to assess forecasters.  Figure \ref{fig:comb_box}e shows that the Brier Score changes minimally with sample size, unlike it's BSC and BSR components. However, the distinction between forecasters is still minimal under this metric. Unlike posterior probability of calibration, it is difficult to interpret an individual forecaster's Brier Score in isolation.

\begin{figure}[h!]
\begin{center}
\includegraphics[width=6.3in]{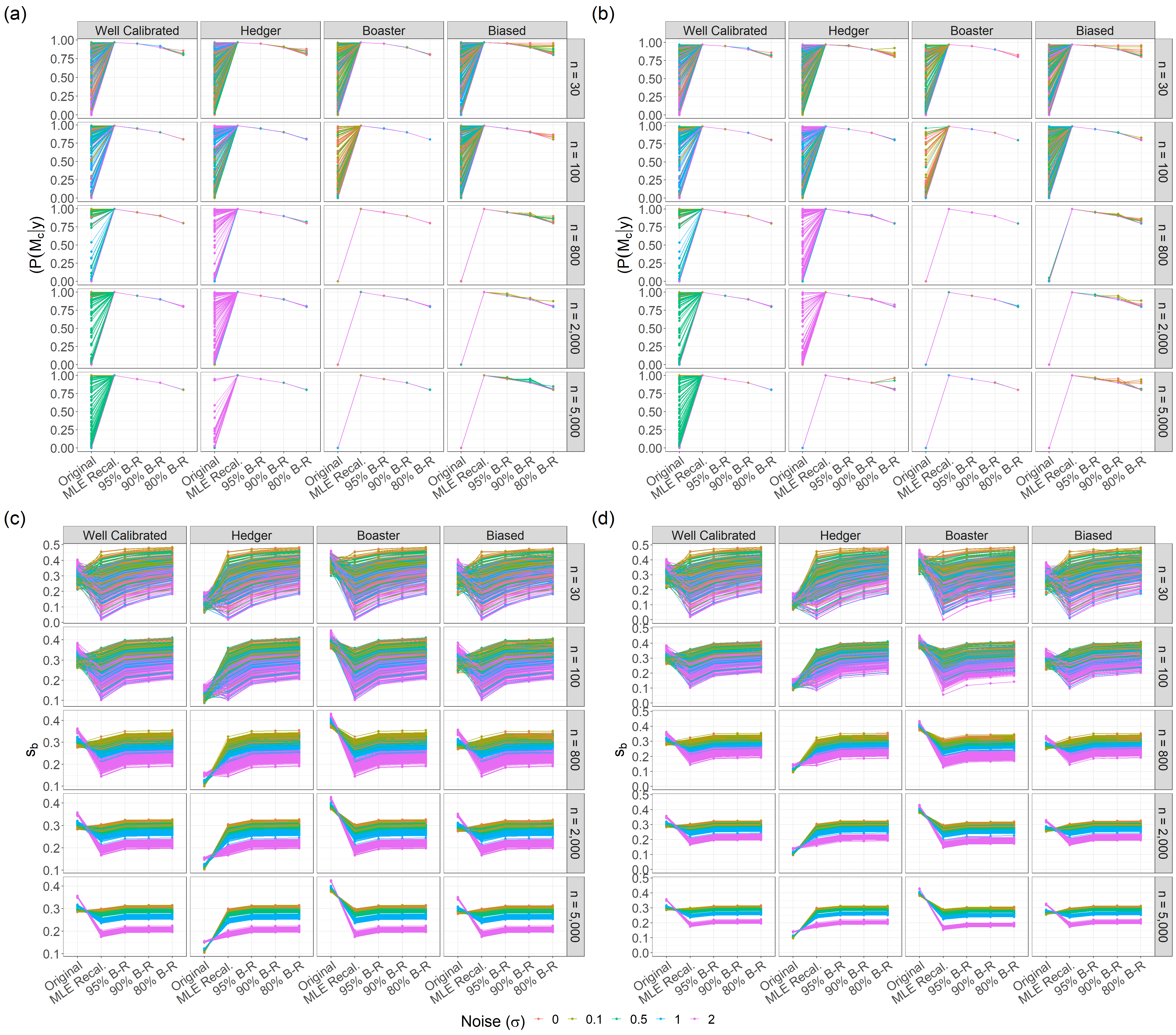}
\vspace{-0.5cm}
\end{center}
\caption{ Lineplots summarizing the change in (a) $P(M_c|\mathbf{y})$ or (c) $s_b$ on the y-axis for 100 MC runs on LLO-miscalibrated simulated forecasters.  Panels (b) and (d) show  $P(M_c|\mathbf{y})$ and $s_b$ under Prelec miscalibration. Sample size increase with vertical panel.  Horizontal panels denote the forecaster type.  The first column of points in each panel represents the value of each metric for the original set. The second, third, fourth, and fifth columns of points represent the same metric for the MLE recalibrated, 95\%, 90\%, and 80\% B-R sets, respectively.  A line is used to connect points that correspond to the same original set of predictions.  The lines and points are colored based on the amount of added noise. Note that the points/lines are plotted in a randomized order. The few sets where boldness-recalibration failed as explained in the main manuscript are removed from this plot. 
\label{fig:lp_comb1}}
\end{figure}

\begin{figure}[h!]
\begin{center}
\includegraphics[width=6.2in]{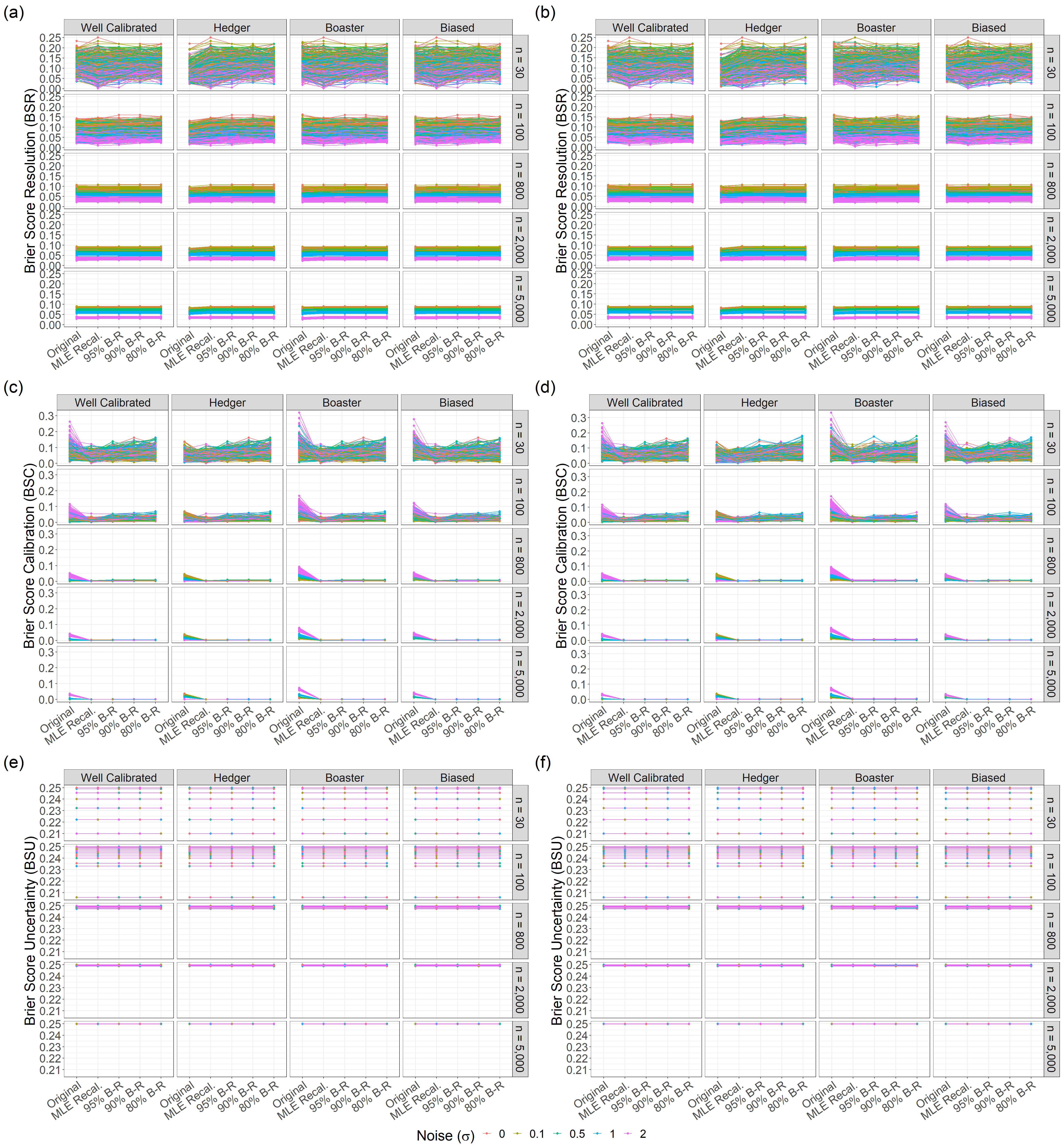}
\vspace{-0.65cm}
\end{center}
\caption{ Lineplots summarizing the change in (a) Brier Score Resolution, (c) Brier Score Calibration, and (e) Brier Score Uncertainty on the y-axis for 100 MC runs on LLO-miscalibrated simulated forecasters.  Panels (b), (d), and (f) show BSR, BSC, and BSU respectively under Prelec miscalibration.  Sample size increase with vertical panel.  Horizontal panels denote the forecaster type.  The first column of points in each panel represents the value of each metric for the original set. The second, third, fourth, and fifth columns of points represent the same metric for the MLE recalibrated, 95\%, 90\%, and 80\% B-R sets, respectively.  A line is used to connect points that correspond to the same original set of predictions.  The lines and points are colored based on the amount of added noise. Note that the points/lines are plotted in a randomized order. The few sets where boldness-recalibration failed as explained in the main manuscript are removed from this plot. 
\label{fig:lp_comb2}}
\end{figure}

\begin{figure}[h!]
\begin{center}
\includegraphics[width=6.3in]{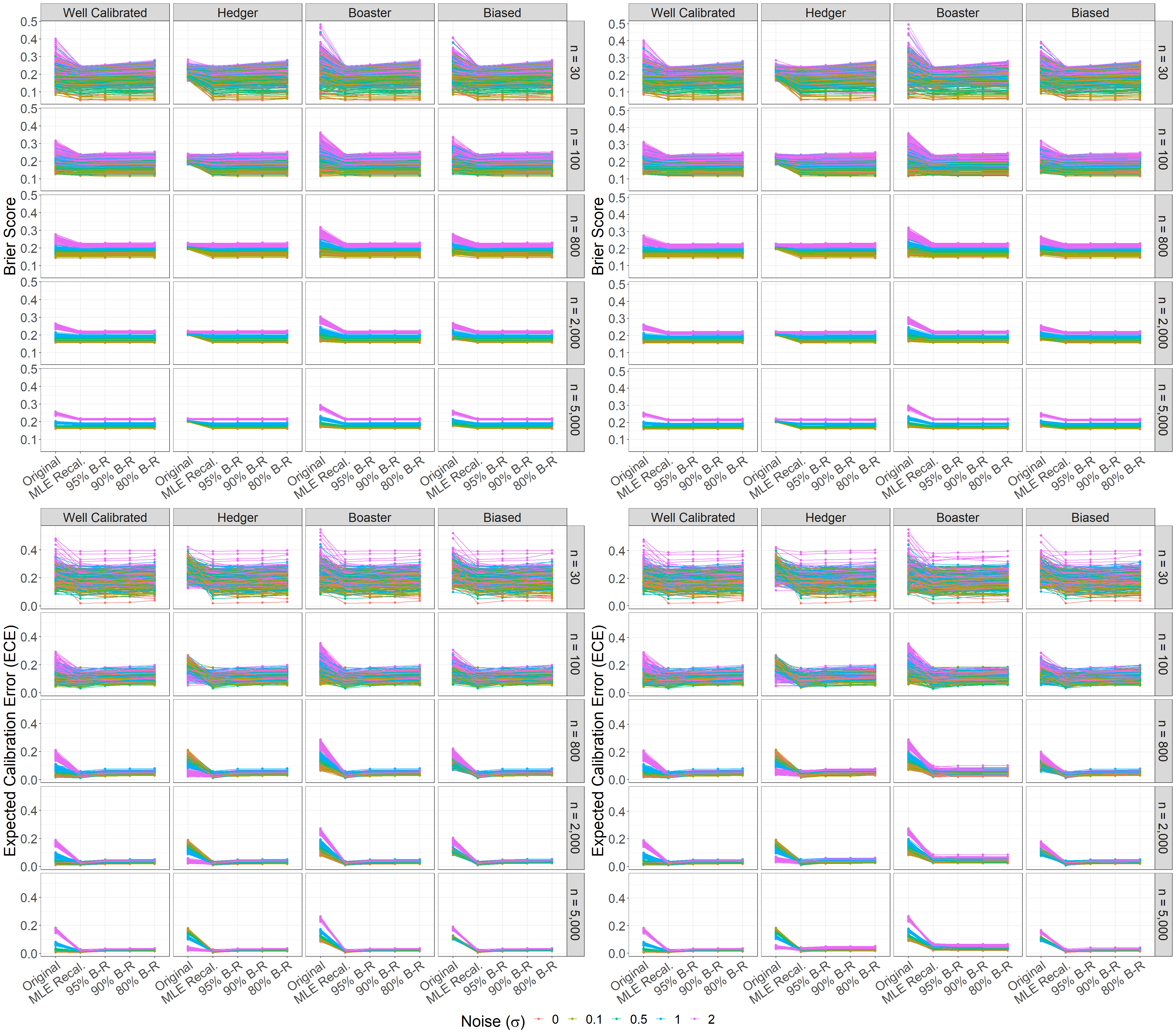}
\vspace{-0.5cm}
\end{center}
\caption{ Lineplots summarizing the change in (a) the overall Brier Score and (c) Expected Calibration Error on the y-axis for 100 MC runs on LLO-miscalibrated simulated forecasters.  Panels (b) and (d) show the Brier Score and ECE under Prelec miscalibration.  Sample size increase with vertical panel.  Horizontal panels denote the forecaster type.  The first column of points in each panel represents the value of each metric for the original set. The second, third, fourth, and fifth columns of points represent the same metric for the MLE recalibrated, 95\%, 90\%, and 80\% B-R sets, respectively.  A line is used to connect points that correspond to the same original set of predictions.  The lines and points are colored based on the amount of added noise. Note that the points/lines are plotted in a randomized order. The few sets where boldness-recalibration failed as explained in the main manuscript are removed from this plot. 
\label{fig:lp_comb3}}
\end{figure}

Figures \ref{fig:lp_comb1}, \ref{fig:lp_comb2}, \ref{fig:lp_comb3}, summarize the change in $P(M_c|\mathbf{y})$, $s_b$,  BSR, BSC, BSU, overall Brier Score, and ECE, respectively, moving from the original simulated set to MLE recalibration to 95\%, 90\% and 80\% boldness-recalibration under both LLO and Prelec Miscalibration. These lineplots show the value of each metric on the y-axis.  Sample size increase with vertical panel.  Horizontal panels denote the forecaster type.  It is important to note the y-axis is not fixed across vertical panels.  The first column of points in each panel represents the value of each metric for the original set of predictions.  MLE recalibration set.  The second column of points represents the same metric for the MLE recalibration set.  The third, fourth, and fifth columns represent the same metric for the 95\%, 90\%, and 80\% B-R sets, respectively.  A line is used to connect points that correspond to the same original set of predictions.  The lines and points are colored based on the amount of added noise.  These results are an extension of those show in the main text.

\subsubsection{Likelihood Ratio Test for Calibration Results}

\begin{figure}[h!]
\begin{center}
\includegraphics[width=4.75in]{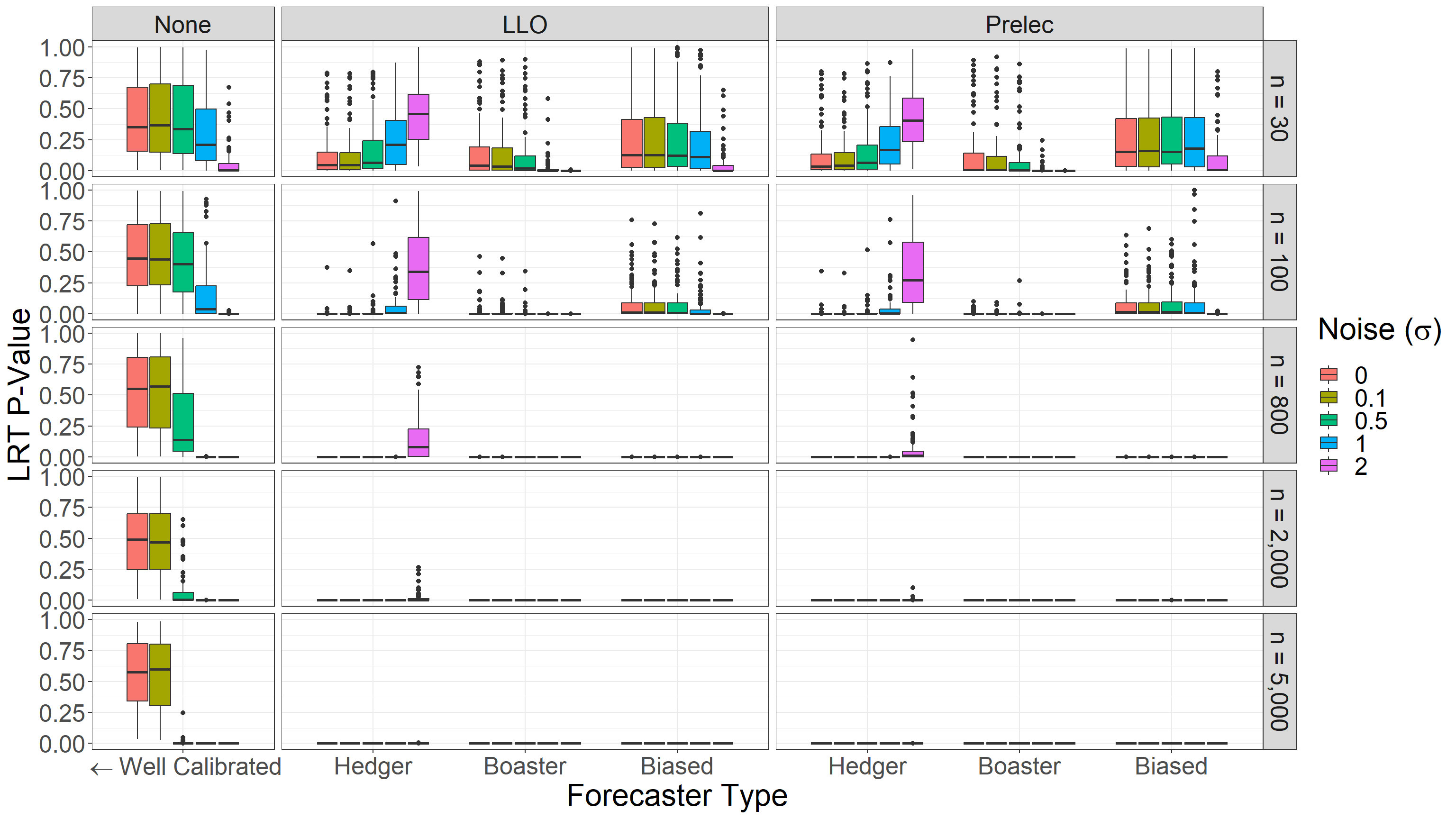}
\vspace{-0.7cm}
\end{center}
\caption{ Boxplots summarizing the P-value from the Likelihood Ratio Test of calibration for 100 MC runs on simulated forecasters.   Boxplots are grouped by forecaster type on the x-axis.  Within groups, added noise increases from left to right.  Only the leftmost boxplot in the Well Calibrated group is perfectly calibrated, and ← indicates calibration increases as noise decreases.  Horizontal panels indicate which adjustment (if any) was applied to create the forecaster type. As sample size increases with vertical panels, the P-value decreases for all forecasters except Well Calibrated with little to no added noise. 
\label{fig:pval_box}}
\vspace{-0.7cm}
\end{figure}

Figure \ref{fig:pval_box} shows that as sample size increases, the LRT p-value for calibration also decreases in all settings except for the Well Calibrated forecasters with little or no noise, where the null hypothesis is true by construction (and thus p-values are uniform).  This reflects increasing power to detect miscalibration with increasing sample size.  Again, notice that Hedgers with large added noise appear well calibrated, but as power increases, our LRT more easily detects the miscalibration.  

\begin{figure}[h!]
\begin{center}
\includegraphics[width=6in]{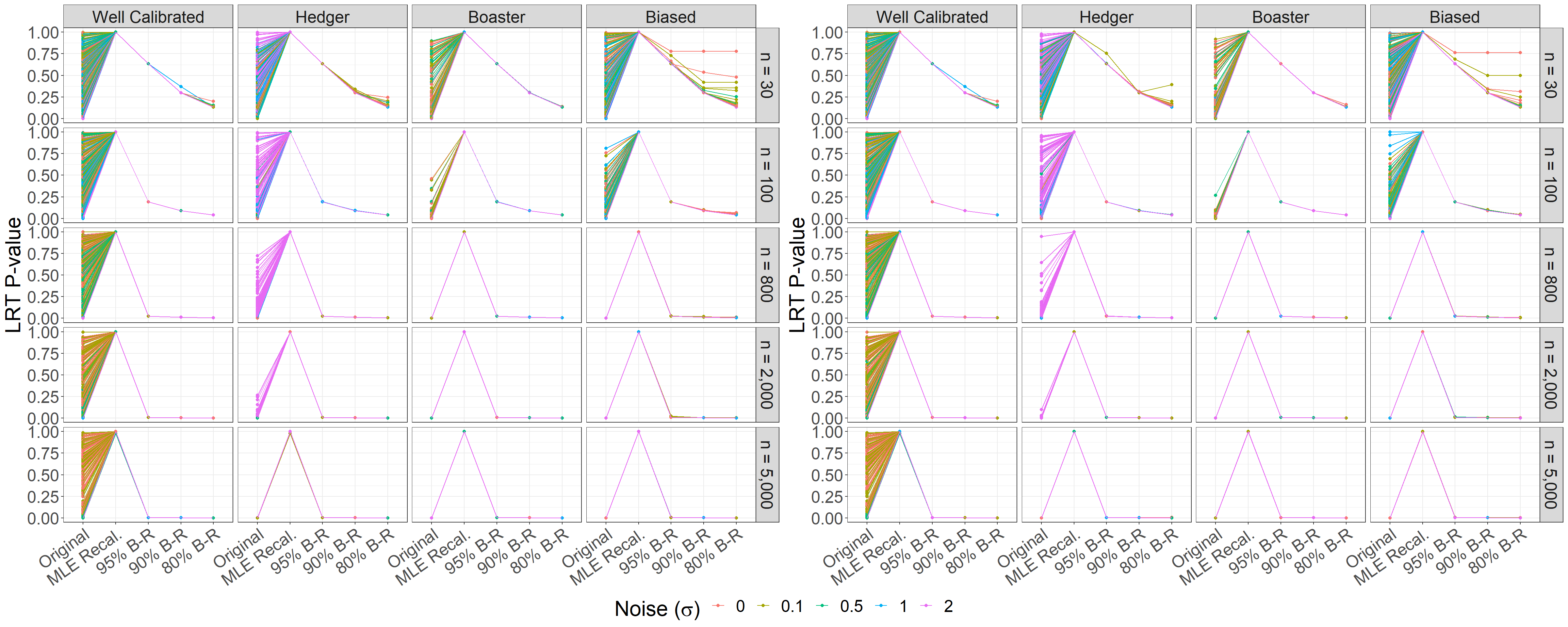}
\vspace{-0.7cm}
\end{center}
\caption{ Lineplots summarizing the change in the P-value from the Likelihood Ratio Test of Calibration on the y-axis for 100 MC runs on LLO-miscalibrated simulated forecasters.  Sample size increase with vertical panel.  Horizontal panels denote the forecaster type.  The first column of points in each panel represents the value of each metric for the original set. The second, third, fourth, and fifth columns of points represent the same metric for the MLE recalibrated, 95\%, 90\%, and 80\% B-R sets, respectively.  A line is used to connect points that correspond to the same original set of predictions.  The lines and points are colored based on the amount of added noise. Note that the points/lines are plotted in a randomized order. The few sets where boldness-recalibration failed as explained in the main manuscript are removed from this plot.
\label{fig:lp_comb4}}
\vspace{-0.75cm}
\end{figure}

Figure \ref{fig:lp_comb4} summarizes the change in the p-value across the original, MLE recalibrated, and 95\%, 90\%, and 80\% boldness-recalibration sets.  Notice that in nearly except Biased forecasters, added noise causes little change in the p-values within each forecaster type after MLE or boldness-recalibration.  At small sample sizes, the LRT would fail to reject the null hypothesis of calibration under all three levels of boldness-recalibration.  However, as sample size increases, the LRT better detects the sacrifice of calibration introduced by boldness-recalibration.  

\end{document}